
%
\font\bigbf=cmbx10  scaled\magstep1

\font\ggreatrm=cmr10  scaled\magstep4

%

%

%

%

%

%

%
%

%

%

%

%
\def\etal{{\it et al.\/}}
\def\tento #1 {\cdot 10^{#1}}
\def\Bigroman#1{\uppercase\expandafter{\romannumeral #1}}
\def\Sggreat{{\hbox{\lower 4pt \hbox{\ggreatrm S}}}}
\def\bvec#1 {{\bf #1}}
\def\simlt{\lower.5ex\hbox{$\; \buildrel < \over \sim \;$}}
\def\simgt{\lower.5ex\hbox{$\; \buildrel > \over \sim \;$}}
\def\rahm #1,#2 {#1^{\hbox{\sevenrm h}}~#2^{\hbox{\sevenrm m}}}
\def\decdm #1,#2 {#1^o~#2'}
%
%
\def\today{%
     \ifcase\month\or January\or Febuary\or March\or April\or May\or
     June\or July\or August\or September\or October\or
     November\or December\fi \enspace\number\day, \enspace\number\year.}
\def\clock{\count0=\time \divide\count0 by 60
    \count1=\count0 \multiply\count1 by -60 \advance\count1 by \time
    \number\count0:\ifnum\count1<10{0\number\count1}\else\number\count1\fi}

\footline={\hss {\rm -- \folio\ --} \hss}


\def\draft#1{\headline={ %
{\bf DRAFT:\ \jobname.tex ~~---~~ version \# #1}
{\sevenrm \hss \today ~~~ Hrs.~\clock}}
  }

%
%

%
\newskip\eightptskip
	\eightptskip = 8pt  plus 1pt minus 1pt

\newskip\sixptskip
	\sixptskip = 6pt  plus 1pt minus 1pt

\def\oneskip{\vskip\baselineskip}

\newskip\hskip
	\hskip = \baselineskip
	\divide\hskip by 2
\newskip\tskip
	\tskip = \baselineskip
	\divide\tskip by 3
\def\halfskip{\vskip \hskip}
\newskip\hoskipup
	\hoskipup = \hskip
	\multiply\hoskipup by -1

\newskip\oskipup
	\oskipup = \hoskipup
	\multiply\hoskipup by 2

\newskip \nullskip
\nullskip = 0pt plus 3 pt minus 3 pt
%
\newskip \ssectskip
	 \ssectskip = \baselineskip
\def\sectvskip{\vskip\ssectskip}
\newskip \ssubsectskip
	 \ssubsectskip = \hskip
\def\subsectvskip{\vskip\ssubsectskip}
\newskip \ssubsubsectskip
	 \ssubsubsectskip = \tskip


%
%

%
\def\pap#1#2\par{ \clabel{#1} \goodbreak \par %
		\noindent \hangindent 2 truecm
		{\hbox to 1 truecm{\hfil [\cref{#1}]}}#2}
\def\ref{\penalty -100 \par \noindent \hangindent \parindent}
\def\sequel{\hbox to 3truecm{\hrulefill}}
\newif\iflongrefs\longrefsfalse
\newif\ifoldrefs\oldrefsfalse

\gdef\RefsDefs{

\iflongrefs

	\immediate\write16{ --> I'm using the reference full format }

	\def\aa ##1 ##2|{{\it Astronomy \& Astrophysics} {\bf ##1}, ##2.}
	\def\araa ##1 ##2|{{\it Annual Review of Astronomy \& %
		Astrophysics} {\bf ##1}, ##2.}
	\def\aj ##1 ##2|{{\it Astronomical Journal} {\bf ##1}, ##2.}
	\def\apj ##1 ##2|{{\it Astrophysical Journal} {\bf ##1}, ##2.}
	\def\apjpress{{\it Astrophysical Journal} in press.}
	\def\apjl ##1 ##2|{{\it  Astrophysical Journal (Letters)} {\bf ##1}, L##2.}
	\def\apjlpress {{\it Astrophysical Journal (Letters)} in press.}
	\def\apjsupp ##1 ##2|{{\it Astrophysical Journal Supplement  %
		Series} {\bf ##1}, ##2.}
	\def\mnras ##1 ##2|{{\it Monthly Notices of the Royal astronomical   %
		Society} {\bf ##1}, ##2.}
	\def\qjras ##1 ##2|{{\it Quarterly Journal of the Royal astronomical   %
		Society} {\bf ##1}, ##2.}
	\def\nat ##1 ##2|{{\it Nature} {\bf ##1}, ##2.}
	\def\pasp ##1 ##2|{{\it Publications of the Astronomical Society of the
Pacific} {\bf ##1}, ##2.}
	\def\physlb ##1 ##2|{{\it Physics Letters B} {\bf ##1}, ##2.}
	\def\physrep ##1 ##2|{{\it Physics Reports} {\bf ##1}, ##2.}
	\def\physrev ##1 ##2|{{\it Physical Review} {\bf ##1}, ##2.}
	\def\physrevb ##1 ##2|{{\it Physical Review B} {\bf ##1}, ##2.}
	\def\physrevd ##1 ##2|{{\it Physical Review D} {\bf ##1}, ##2.}
	\def\physrevl ##1 ##2|{{\it Physical Review Letters} {\bf ##1}, ##2.}
	\def\sovastr ##1 ##2|{{\it Soviet Astronomy} {\bf ##1}, ##2.}
	\def\sovastrl ##1 ##2|{{\it Soviet Astronomy (Letters)} {\bf ##1}, L##2.}
	\def\commastr ##1 ##2|{{\it Comments Astrophys.} {\bf ##1}, ##2.}
	\def\book ##1 {{\it ``{##1},''\ }}

\else

	\ifoldrefs

		\immediate\write16{ --> I'm using the old reference short format }
		\def\aa ##1 ##2|{{\it Astr. \& Astrophys.} {\bf ##1}, ##2.}
		\def\araa ##1 ##2|{{\it Ann. Rev. Astr. \& Astrophys.} {\bf ##1}, ##2.}
		\def\aj ##1 ##2|{{\it A.J.} {\bf ##1}, ##2.}
		\def\apj ##1 ##2|{{\it Ap.J.} {\bf ##1}, ##2.}
		\def\apjpress{{\it Ap.J.} in press.}
		\def\apjl ##1 ##2|{{\it  Ap.J. (Lett.)} {\bf ##1}, L##2.}
		\def\apjlpress{{\it  Ap.J. (Lett.)} in press.}
		\def\apjsupp ##1 ##2|{{\it Ap.J. Supp.} {\bf ##1}, ##2.}
		\def\mnras ##1 ##2|{{\it M.N.R.\-A.S.} {\bf ##1}, ##2.}
		\def\mnraspress{{\it M.N.R.\-A.S.} in press.}
		\def\qjras ##1 ##2|{{\it Q.J.R.\-a.S.} {\bf ##1}, ##2.}
		\def\nat ##1 ##2|{{\it Nat.} {\bf ##1}, ##2.}
		\def\pasp ##1 ##2|{{\it Pub. Astr. Soc. Pac.} {\bf ##1}, ##2}
		\def\physlb ##1 ##2|{{\it Phys. Lett. B} {\bf ##1}, ##2.}
		\def\physrep ##1 ##2|{{\it Phys. Rep.} {\bf ##1}, ##2.}
	       	\def\physrev ##1 ##2|{{\it Phys. Rev.} {\bf ##1}, ##2.}
		\def\physrevb ##1 ##2|{{\it Phys. Rev. B} {\bf ##1}, ##2.}
		\def\physrevd ##1 ##2|{{\it Phys. Rev. D} {\bf ##1}, ##2.}
		\def\physrevl ##1 ##2|{{\it Phys. Rev. Lett.} {\bf ##1}, ##2.}
		\def\sovastr ##1 ##2|{{\it Sov. Astr.} {\bf ##1}, ##2.}
		\def\sovastrl ##1 ##2|{{\it Sov.  Astr. (Lett.)} {\bf ##1}, L##2.}
		\def\commastr ##1 ##2|{{\it Comm. Astr.} {\bf ##1}, ##2.}
		\def\book ##1 {{\it ``{##1}''\ }}

	\else

		\immediate\write16{ --> I'm using the new reference short format }

		\def\aa ##1 ##2|{ A\&A {##1}, ##2}
		\def\araa ##1 ##2|{ ARA\&A {##1}, ##2}
		\def\aj ##1 ##2|{ AJ {##1}, ##2}
		\def\apj ##1 ##2|{ ApJ {##1}, ##2}
		\def\apjpress{ ApJ in press}
		\def\apjl ##1 ##2|{ ApJ {##1}, L##2}
		\def\apjlpress{ ApJ in press}
		\def\apjsupp ##1 ##2|{ ApJS {##1}, ##2}
		\def\mnras ##1 ##2|{ MNR\-AS {##1}, ##2}
		\def\mnraspress{ MNR\-AS in press}
		\def\qjras ##1 ##2|{ QJR\-AS { ##1}, ##2}
		\def\nat ##1 ##2|{ Nat {##1}, ##2}
		\def\pasp ##1 ##2|{{ PASP} {##1}, ##2}
		\def\physlb ##1 ##2|{ Phys Lett B{##1}, ##2}
		\def\physrep ##1 ##2|{ Phys Rep {##1}, ##2}
		\def\physrev ##1 ##2|{ Phys Rev {##1}, ##2}
		\def\physrevb ##1 ##2|{ Phys Rev B {##1}, ##2}
		\def\physrevd ##1 ##2|{ Phys Rev D {##1}, ##2}
		\def\physrevl ##1 ##2|{ Phys Rev Lett {##1}, ##2}
		\def\sovastr ##1 ##2|{ SvA {##1}, ##2}
		\def\sovastrl ##1 ##2|{ SvA {##1}, L##2}
		\def\commastr ##1 ##2|{ Comm Astr {##1}, ##2}
		\def\book ##1 { {##1} }

	\fi

	\def\bookit ##1 {{\it ``{##1}''}}

	\def\replica{\kern \parindent \hbox to 1 truecm{\hrulefill}\ }

\fi

}
%

%

\def\doublencite#1 #2{$^{\,\cref{#1},\cref{#2}}$}

\def\fromtoncite#1 #2{$^{\,\cref{#1}-\cref{#2}}$}

\def\nref#1{\penalty -100 \par \noindent \hangindent 1 truein 
\hbox to 0.5 truein {\hfil [\cref{#1}]}~}

\newif\ifapjstyle

\newif\ifapjnumbering
\def\apjnumbering{
	\nopagenumbers
	\headline={\ifnum\pageno > 1
			\hfil \folio \hfil
		   \else
			\hfil
			\fi}
                  }

\newif\ifsimboli
\newif\ifriferimenti


\newif\ifsimboli
\newif\ifriferimenti

\newwrite\filerefs
\newwrite\fileeqs
\def\simboli{
    \immediate\write16{ !!! Genera il file \jobname.REFS }
    \simbolitrue\immediate\openout\filerefs=\jobname.refs
    \immediate\write16{ !!! Genera il file \jobname.EQS }
    \simbolitrue\immediate\openout\fileeqs=\jobname.eqs}

\newwrite\fileausiliario
\def\riferimenti{
    \riferimentitrue\openin1 \jobname.aux
    \ifeof1\relax\else\closein1\relax\input\jobname.aux
    \immediate\write16{ --> Legge il file \jobname.AUX }  \fi
    \immediate\openout\fileausiliario=\jobname.aux
    \immediate\write16{ !!! Genera il file \jobname.AUX } }


%
%

\newcount\chapnum\global\chapnum=0
\newcount\sectnum\global\sectnum=0
\newcount\subsectnum\global\subsectnum=0
\newcount\subsubsectnum\global\subsubsectnum=0
\newcount\eqnum\global\eqnum=0
\newcount\citnum\global\citnum=0
\newcount\fignum\global\fignum=0
\newcount\tabnum\global\tabnum=0

\newcount\nnummore\global\nnummore=0

\def\itemore{\advance \nnummore by 1 \item{[\the\nnummore]}}

\newcount\nnummorerm\global\nnummorerm=0

\def\itemorerm{\advance \nnummorerm by 1 %
	\item{\expandafter{\romannumeral\the\nnummorerm})} }

\def\thechaproman{\uppercase\expandafter{\romannumeral \the \chapnum}}

\def\Chap#1{Chap.~\uppercase\expandafter{\romannumeral #1}}

\def\Figcapt#1#2{ \dimen150= \hsize  \dimen151= \hsize \dimen152= 2.25 truecm%
		\advance \dimen150 by -2pt \relax %
		\advance \dimen151 by -\dimen152 \relax %
		\advance \dimen151 by -4pt  \relax %
		\setbox151  \hbox{\hsize = \dimen151 %
			\noindent \vtop{\noindent #2}}
		\setbox150 \hbox{\hsize = \dimen150 %
     			\noindent \hbox{\hsize = \dimen151 %
				\bf Figure~\figref{#1}: \hfil}%
		\hfil\box151}
		\noindent\box150}

\def\Tabcapt#1#2{ \dimen150= \hsize  \dimen151= \hsize \dimen152= 2.25 truecm%
		\advance \dimen150 by -2pt \relax %
		\advance \dimen151 by -\dimen152 \relax %
		\advance \dimen151 by -4pt  \relax %
		\setbox151  \hbox{\hsize = \dimen151 %
			\noindent \vtop{\noindent #2}}
		\setbox150 \hbox{\hsize = \dimen150 %
     			\noindent \hbox{\hsize = \dimen151 %
				\bf Table~\figref{#1}: \hfil}%
		\hfil\box151}
		\noindent\box150}

%
%
%

\newif\ifndoppia
\def\numerazionedoppia{\ndoppiatrue\gdef\lasezionecorrente{
\the\sectnum }}

\def\seindefinito#1{\expandafter\ifx\csname#1\endcsname\relax}
\def\spoglia#1>{}

\def\cref#1{\seindefinito{@c@#1}\immediate\write16{ !!! \string\cref{#1}
    non definita !!!}
    \expandafter\xdef\csname@c@#1\endcsname{??}\fi\csname@c@#1\endcsname}

\def\eqref#1{\seindefinito{@eq@#1}\immediate\write16{ !!! \string\eqref{#1}
    non definita !!!}
    \expandafter\xdef\csname@eq@#1\endcsname{??}\fi\csname@eq@#1\endcsname}

\def\sectref#1{\seindefinito{@s@#1}\immediate\write16{ !!! \string\sectref{#1}
    non definita !!!}
    \expandafter\xdef\csname@s@#1\endcsname{??}\fi\csname@s@#1\endcsname}

\def\figref#1{\seindefinito{@f@#1}\immediate\write16{ !!! \string\figref{#1}
    non definita !!!}
    \expandafter\xdef\csname@f@#1\endcsname{??}\fi\csname@f@#1\endcsname}

\def\tabref#1{\seindefinito{@f@#1}\immediate\write16{ !!! \string\tabref{#1}
    non definita !!!}
    \expandafter\xdef\csname@f@#1\endcsname{??}\fi\csname@f@#1\endcsname}

%

\def\section#1\par{\immediate\write16{#1}\goodbreak\oneskip\halfskip
    \noindent{\bigbf #1}\nobreak\oneskip \nobreak\noindent}

\def\autosection#1#2\par { %
    \global\advance\sectnum by 1   %
 	\global\subsectnum=0
	\global\nnummore=0
	\global\nnummorerm=0
	\ifndoppia
		\global\eqnum=0
		\global\fignum=0
	    	\global\tabnum=0
		\fi
    \xdef\lasezionecorrente{\the\sectnum}
    \def\usaegetta{1}\seindefinito{@s@#1}\def\usaegetta{2}\fi

\expandafter\ifx\csname@s@#1\endcsname\lasezionecorrente\def\usaegetta{2}\fi
    \ifodd\usaegetta\immediate\write16
      { !!! possibili riferimenti errati a \string\sectref{#1} }\fi
    \expandafter\xdef\csname@s@#1\endcsname{\lasezionecorrente}
    \immediate\write16{\lasezionecorrente.#2}
    \ifsimboli
      \immediate\write\filerefs{ }\immediate\write\filerefs{ }
      \immediate\write\filerefs{  Sezione \lasezionecorrente : %
 sectref.   #1         page: \the\pageno}
      \immediate\write\filerefs{ }
      \immediate\write\fileeqs{ }\immediate\write\fileeqs{ }
      \immediate\write\fileeqs{  Sezione \lasezionecorrente : %
 sectref.   #1         page: \the\pageno}
      \immediate\write\fileeqs{ } \fi
    \ifriferimenti
      \immediate\write\fileausiliario{\string\expandafter\string\edef
      \string\csname@s@#1\string\endcsname{\lasezionecorrente}}\fi
\goodbreak \sectvskip %
\vtop{ %
\begingroup \hangindent 0.505truecm
	\noindent \hbox to  0.50truecm {{\bigbf\lasezionecorrente} \hfill}%
	{\bigbf #2} \par%
\endgroup
	}
	  \nobreak \vskip -\parskip \nobreak\noindent 
	} %

\def\autosubsection#1#2\par { %
    \global\advance\subsectnum by 1
			\ifndoppia
				\global\eqnum=0
				\global\fignum=0
			    	\global\tabnum=0
			    	\global\subsubsectnum=0
				\global\nnummore=0
				\global\nnummorerm=0
				\fi
    \xdef\lasezionecorrente{\the\sectnum.\the\subsectnum}
    \def\usaegetta{1}\seindefinito{@s@#1}\def\usaegetta{2}\fi

\expandafter\ifx\csname@s@#1\endcsname\lasezionecorrente\def\usaegetta{2}\fi
    \ifodd\usaegetta\immediate\write16
      { !!! possibili riferimenti errati a \string\sectref{#1} }\fi
    \expandafter\xdef\csname@s@#1\endcsname{\lasezionecorrente}
    \immediate\write16{\lasezionecorrente.#2}
    \ifsimboli
      \immediate\write\filerefs{ } %
      \immediate\write\filerefs{  Sezione \lasezionecorrente : %
 sectref.   #1         page: \the\pageno}
      \immediate\write\filerefs{ }
      \immediate\write\fileeqs{ }  %
      \immediate\write\fileeqs{  Sezione \lasezionecorrente : %
 sectref.   #1         page: \the\pageno}
      \immediate\write\fileeqs{ } \fi
    \ifriferimenti
      \immediate\write\fileausiliario{\string\expandafter\string\edef
      \string\csname@s@#1\string\endcsname{\lasezionecorrente}}\fi
\goodbreak   
	\subsectvskip %
\vtop{ %
	\begingroup \hangindent 0.705truecm
	\noindent \hbox to  0.70truecm {{\it\lasezionecorrente} \hfill}%
	{\it #2} \par%
	\endgroup
	}
  \nobreak \vskip -\parskip \nobreak\noindent
	} 

\def\autosubsubsection#1#2\par { %
    \global\advance\subsubsectnum by 1
				\global\nnummore=0
				\global\nnummorerm=0
    \xdef\lasezionecorrente{\the\sectnum.\the\subsectnum.\the\subsubsectnum}
    \def\usaegetta{1}\seindefinito{@s@#1}\def\usaegetta{2}\fi

\expandafter\ifx\csname@s@#1\endcsname\lasezionecorrente\def\usaegetta{2}\fi
    \ifodd\usaegetta\immediate\write16
      { !!! possibili riferimenti errati a \string\sectref{#1} }\fi
    \expandafter\xdef\csname@s@#1\endcsname{\lasezionecorrente}
    \immediate\write16{\lasezionecorrente.#2}
    \ifsimboli
      \immediate\write\filerefs{ }\immediate\write\filerefs{ }
      \immediate\write\filerefs{  Sezione \lasezionecorrente : %
 sectref.   #1         page: \the\pageno}
      \immediate\write\filerefs{ }
      \immediate\write\fileeqs{ }\immediate\write\fileeqs{ }
      \immediate\write\fileeqs{  Sezione \lasezionecorrente : %
 sectref.   #1         page: \the\pageno}
      \immediate\write\fileeqs{ } \fi
    \ifriferimenti
      \immediate\write\fileausiliario{\string\expandafter\string\edef
      \string\csname@s@#1\string\endcsname{\lasezionecorrente}}\fi
    \goodbreak  
	\subsectvskip  %
	\noindent \item{{\it\lasezionecorrente}}%
	{\it #2} \par %
	  \nobreak\halfskip \vskip -\parskip \nobreak\noindent
	} 

\def\semiautosection#1#2\par{
    \gdef\lasezionecorrente{#1}\ifndoppia\global\eqnum=0\fi
    \ifsimboli
      \immediate\write\filesimboli{ }\immediate\write\filesimboli{ }
      \immediate\write\filesimboli{  Sezione ** : sref.
          \expandafter\spoglia\meaning\lasezionecorrente}
      \immediate\write\filesimboli{ }\fi
    \section#2\par}

\def\eqlabel#1{\global\advance\eqnum by 1
    \ifndoppia\xdef\ilnumero{\lasezionecorrente.\the\eqnum}
       \else\xdef\ilnumero{\the\eqnum}\fi
    \def\usaegetta{1}\seindefinito{@eq@#1}\def\usaegetta{2}\fi
    \expandafter\ifx\csname@eq@#1\endcsname\ilnumero\def\usaegetta{2}\fi
    \ifodd\usaegetta\immediate\write16
       { !!! possibili riferimenti errati a \string\eqref{#1} }\fi
    \expandafter\xdef\csname@eq@#1\endcsname{\ilnumero}
    \ifndoppia
       \def\usaegetta{\expandafter\spoglia\meaning %
			\lasezionecorrente.\the\eqnum}
       \else\def\usaegetta{\the\eqnum}\fi
    \ifsimboli
       \immediate\write\fileeqs{     Equazione \ilnumero :%
  eqref.   #1           page: \the\pageno}\fi
    \ifriferimenti
       \immediate\write\fileausiliario{\string\expandafter\string\edef
       \string\csname@eq@#1\string\endcsname{\usaegetta}}\fi}

\def\autoeqno#1{\eqlabel{#1}
	\expandafter\eqno \hbox{%
	({\rm\ilnumero\kern 0.1ex})}
	}

\def\autoleqno#1{\eqlabel{#1}\leqno(\hbox{\rm \csname@eq@#1\endcsname})}

\def\eqmore{\global\advance\eqnum by 1
    \ifndoppia\xdef\ilnumero{\lasezionecorrente.\the\eqnum}
       \else\xdef\ilnumero{\the\eqnum}\fi
	\expandafter\eqno \hbox{%
	({\rm\ilnumero\kern 0.1ex})}
	}

\def\figlabel#1{\global\advance\fignum by 1
    \ifndoppia\xdef\ilnumero{\lasezionecorrente.\the\eqnum}
       \else\xdef\ilnumero{\the\fignum}\fi
    \def\usaegetta{1}\seindefinito{@f@#1}\def\usaegetta{2}\fi
    \expandafter\ifx\csname@f@#1\endcsname\ilnumero\def\usaegetta{2}\fi
    \ifodd\usaegetta\immediate\write16
       { !!! possibili riferimenti errati a \string\figref{#1} }\fi
    \expandafter\xdef\csname@f@#1\endcsname{\ilnumero}
    \ifndoppia
       \def\usaegetta{\expandafter\spoglia\meaning %
			\lasezionecorrente.\the\fignum}
       \else\def\usaegetta{\the\fignum}\fi
    \ifsimboli
       \immediate\write\fileeqs{                    Figure \figref{#1} : %
 figref.   #1               page: \the\pageno}\fi
    \ifriferimenti
       \immediate\write\fileausiliario{\string\expandafter\string\edef
       \string\csname@f@#1\string\endcsname{\usaegetta}}\fi}

%
%

\def\tablabel#1{\global\advance\tabnum by 1
    \ifndoppia\xdef\ilnumero{\lasezionecorrente.\the\tabnum}
       \else\xdef\ilnumero{\the\tabnum}\fi
    \def\usaegetta{1}\seindefinito{@f@#1}\def\usaegetta{2}\fi
    \expandafter\ifx\csname@f@#1\endcsname\ilnumero\def\usaegetta{2}\fi
    \ifodd\usaegetta\immediate\write16
       { !!! possibili riferimenti errati a \string\tabref{#1} }\fi
    \expandafter\xdef\csname@f@#1\endcsname{\ilnumero}
    \ifndoppia
       \def\usaegetta{\expandafter\spoglia\meaning  %
				\lasezionecorrente .\the\tabnum}
       \else\def\usaegetta{\the\tabnum}\fi
    \ifsimboli
       \immediate\write\fileeqs{                    Table \tabref{#1} : %
 tabref.   #1               page: \the\pageno}\fi
    \ifriferimenti
       \immediate\write\fileausiliario{\string\expandafter\string\edef
       \string\csname@f@#1\string\endcsname{\usaegetta}}\fi}

\def\clabel#1{\global\advance\citnum by 1%
  \xdef\lacitazione{\the\citnum}%
  \def\usaegetta{1}\seindefinito{@c@#1}\def\usaegetta{2}\fi%
  \expandafter\ifx\csname@c@#1\endcsname\lacitazione\def\usaegetta{2}\fi%
  \ifodd\usaegetta\immediate\write16%
       { !!! possibili riferimenti errati a \string\cref{#1} }\fi%
\expandafter\xdef\csname@c@#1\endcsname{\lacitazione}%
\ifsimboli%
  \immediate%
  \write\filerefs{   Citazione \lacitazione: #1  page: \the\pageno}\fi%
\ifriferimenti\immediate\write\fileausiliario%
    {\string\expandafter\string\edef\string\csname@c@#1\string\endcsname%
    {\lacitazione}}\fi%
 }%
%
%
\catcode`@ = 11
\newif\if@TestSubString
\def\IfSubString #1#2{%
	\edef\@MainString{#1}%
	\def\@TestSubS ##1#2##2\@Del{\edef\@TestTemp{##1}}%
		\expandafter\@TestSubS \@MainString#2\@Del
		\ifx\@MainString\@TestTemp
			\@TestSubStringfalse
		\else
			\@TestSubStringtrue
		\fi
		\if@TestSubString
		}
\def\Substitute@etal #1etal#2\@Del{\def\dummystring{#1\etal#2}}
\def\checketal#1{%
		\def\dummystring{#1}
		\IfSubString{#1}{etal}%
			\expandafter\Substitute@etal\dummystring\@Del%
		\fi%
\dummystring%
}
\catcode`@ = 12
\def\back3pt{\hbox{\kern -3pt}}
\def\cite #1/{\clabel{#1}\back3pt\checketal{#1}}
%
%
\catcode`@=11

\newdimen\@StrutSkip

\newdimen\@StrutSkipTemp

%
%
%
\def\SetStrut{%
   \@StrutSkip = \baselineskip
   \ifdim\baselineskip < 0pt
    \errhelp = {You probably called \string\offinterlineskip
		before \string\SetStrut}
     \errmessage{\string\SetStrut: negative \string\baselineskip
			(\the\baselineskip)}%
      \fi
        }

%
%
\def\MyStrut{%
	\vrule height 0.7\@StrutSkip
	        depth 0.3\@StrutSkip
 	        width 0pt
             }
%
%
%
\def \HigherStrut #1{%
     \@StrutSkipTemp = 0.7 \@StrutSkip
	\advance \@StrutSkipTemp by #1%
	\vrule height \@StrutSkipTemp depth 0.3\@StrutSkip width 0pt
      }

%
%
%
\def \DeeperStrut #1{%
     \@StrutSkipTemp = 0.3 \@StrutSkip
	\advance \@StrutSkipTemp by #1%
	\vrule height 0.7\@StrutSkip depth \@StrutSkipTemp width 0pt
      }
%
%
%
\def \LargerStrut #1{%
     \@StrutSkipTemp =  \@StrutSkip
	\advance \@StrutSkipTemp by #1%
	\vrule height 0.7\@StrutSkipTemp depth 0.3\@StrutSkipTemp width 0pt
      }

%
%
\SetStrut

%
%
%


\newcount\mscount
\def\multispan #1{%
      \omit
      \mscount = #1 %
      \loop \ifnum \mscount > 1
		\sp@n
	  \repeat
       }
\def\sp@n{\span\omit\advance\mscount by -1}

\catcode`@=12

%
%
%
%
\def\boxit#1{\leavevmode\hbox{\vrule\vtop{\vbox{\kern.33333pt\hrule
    \kern10pt\hbox{\kern10pt\vbox{#1}\kern10pt}}\kern10pt\hrule}\vrule}}

\RefsDefs
\overfullrule 0 pt
\magnification 1200
\def\parn{\par\noindent}
\baselineskip 7 true mm
\hsize 17 true cm

{\centerline {\bf ON THE LUMINOSITY FUNCTION OF EARLY--TYPE GALAXIES }}
\bigskip
Elena ZUCCA$^{(1,2)}$, Lucia POZZETTI$^{(2)}$ and Giovanni ZAMORANI$^{(3,1)}$
\medskip\noindent
$^{(1)}$ Istituto di Radioastronomia del CNR, via Gobetti 101, I-40129 Bologna
\smallskip\noindent
$^{(2)}$ Dipartimento di Astronomia, Universit\`a di Bologna, via Zamboni 33,
I--40126 Bologna
\smallskip\noindent
$^{(3)}$ Osservatorio Astronomico di Bologna, via Zamboni 33, I-40126 Bologna
\vskip 1 true cm
\par
accepted for publication on {\it MNRAS}
\vskip 2 true cm
{\centerline {\bf ABSTRACT}}
\medskip
\par
In a recent paper Loveday et al. (1992) have presented new results
on the luminosity function for a sample of galaxies with $b_J \le 17.15$.
After having morphologically classified each galaxy (early--type, late--type,
merged or uncertain), they have estimated the parameters of a Schechter
luminosity function for early-- and late--type galaxies.
However, in their sample there is a bias against identifying early--type
galaxies at large distances and/or faint magnitudes: in fact, many of the
early--type galaxies at faint magnitudes have probably been classified as
``uncertain". As discussed in Loveday et al., the existence of such a bias
is indicated by the fact that for these galaxies $<V/V_{max}>=0.32$.
\par
In this paper we show, both theoretically and through the use
of simulated samples, that this incompleteness strongly biases the derived
parameters of the luminosity function for early--type galaxies. If no
correction for such incompleteness is applied to the data (as done by
Loveday et al.), one obtains a flatter slope $\alpha$ and a brighter
$M^*$ with respect to the real parameters.
\bigskip
\parn
\vskip 2 true cm
\parn
e--mail: 38057::ZUCCA or zucca@astbo1.bo.cnr.it
\vfill\eject
\parn
{\bf 1. Introduction}
\medskip
\par
An accurate knowledge of the optical luminosity function of galaxies
is required for many applications in cosmology. For instance, it is essential
in interpreting galaxy
number counts and in analyzing the spatial distribution of galaxies from
redshift surveys; in addition, the shape of this function is of theoretical
interest as it may provide constraints on models of galaxy formation.
\par
An interesting question about the luminosity function
of galaxies concerns its universality:
indeed, Binggeli, Sandage \& Tammann (1988) have shown that the luminosity
function depends on the morphological type, particularly at the faint end.
On the other hand, it has been demonstrated that the mix of morphological types
is closely related to the local matter density (Dressler 1980).
The accurate knowledge of the luminosity function for each morphological type
is of great interest also for the models of number--magnitude counts.
These models strongly depend on the morphological mix and therefore
need the knowledge not only of the fraction of the various galaxy
types but also of their K--corrections and luminosity functions.
\par
Loveday et al. (1992) have recently presented new results
on the luminosity function
for a sample of galaxies with $b_J \le 17.15$. After having
morphologically classified each galaxy (early--type, late--type,
merged or uncertain),
they have estimated the parameters of a Schechter luminosity function for
early-- and late--type galaxies, using the STY parametric maximum likelihood
method (Sandage, Tammann \& Yahil 1979). While for late--type galaxies
their parameters are in reasonable agreement with those derived from other
samples (see {\it f.i.} Efstathiou, Ellis \& Peterson 1988), the parameters
for early--type galaxies are not consistent with previous determinations.
\par
As mentioned by Loveday et al., in their sample there is a bias
against identifying early--type galaxies at large distances and/or faint
magnitudes: in fact, many of the early--type galaxies at faint magnitudes
have probably been classified as ``uncertain'', and therefore have not
been used in computing the luminosity function. The existence of such a bias
is demonstrated by the fact that for these galaxies $<V/V_{max}>=0.32$. The
same bias appears not to be present in the classification of the late--type
galaxies, for which $<V/V_{max}>=0.47$ (see Table 1 in Loveday et al. 1992).
In this paper we show, both theoretically and through the use
of simulated samples, that this incompleteness strongly biases the derived
parameters of the luminosity function for early--type galaxies.
\par
In Sect. 2 we demonstrate that the classification incompleteness biases the
results of the STY method and in Sect. 3 we estimate the amount of this bias
through simulations.
\bigskip
\parn
{\bf 2. The luminosity function of galaxies}
\medskip
\par
The luminosity function of galaxies is well represented by a Schechter (1976)
form
$$ \phi(L) dL = \phi^* \left({L\over{L^*}}\right)^{\alpha}
    {\rm e}^{-L/L^*} d \left({L\over{L^*}}\right)     \eqno(1)  $$
where $\alpha$ and $L^*$ are parameters referring to the shape of the function
and $\phi^*$ contains the information about the normalization; these parameters
have to be determined from the data.
\par
Many different methods have been used in the past years to compute
the parameters of the galaxy luminosity function. Recently, however, the STY
method (Sandage et al. 1979) has been the most widely used, and
it has been shown that this estimator is unbiased with respect to density
inhomogeneities (see {\it f.i.} Efstathiou et al. 1988).
The basic idea of this method is to
compute the estimator of the quantity $\displaystyle{ {{\phi}\over{\Phi}} }$,
where $\Phi$ is the integrated luminosity function. Under the assumption that
$\phi(L)$ is not a function of position [{\it i.e.} $ \phi(L, x, y, z)\ dL\ dV
= \rho(x, y, z) dV\ \psi(L) dL $], the probability of seeing a galaxy of
luminosity $L_i$ at redshift $z_i$ is
$$ p_i =  { {\psi(L_i)}\over
{\displaystyle{\int_{L_{min}(z_i)}^\infty \psi(L) dL }}}    \eqno(2) $$
where $L_{min}(z_i)$ is the minimum luminosity observable at redshift $z_i$ in
a magnitude--limited sample.
\par
The best parameters $\alpha$ and $L^*$ of the luminosity function
are then determined by maximizing the likelihood function
${\cal L} (\alpha, L^*)$, which is
the product over all the galaxies of the individual probabilities $p_i$.
This corresponds to minimize the function
$${\cal S}= -2 \ln {\cal L} =     $$
$$   = -2 \left[ \alpha \sum_{i=1}^N \ln L_i
    - N (\alpha + 1) \ln L^* - {1 \over {L^*}}  \sum_{i=1}^N L_i -
     \sum_{i=1}^N \ln \Gamma\left(\alpha+1, {{L_{min}(z_i)}\over{L^*}} \right)
       \right]    \eqno(3) $$
where $\Gamma$ is the incomplete Euler gamma function and $N$ is the
total number of galaxies in the sample.
\par
This formula is correct only for a complete, unbiased sample in which
all galaxies with $m < m_{lim}$ are members of the sample or all galaxies
with $m < m_{lim}$ have the same probability of being members of the sample
(as, for example, in a redshift survey with $\displaystyle{1 \over n}$
sampling). In other cases in which each galaxy of the sample has a different
weight $w_i$, which may be a function of an intrinsic property
of the galaxy ({\it f.i.} the distance, the absolute or apparent
magnitude, the diameter, etc.), eq. (2) is not valid anymore. If we
define the weight $w_i$ as the inverse of the probability that
the $i^{th}$ galaxy has of being included in the sample,
eqs. (2) and (3) have to be modified in:
$$ p_i = \left( { {\psi(L_i)}\over
{\displaystyle{\int_{L_{min}(z_i)}^\infty \psi(L) dL }}} \right)^{w_i}
   \eqno(4) $$
and
$$  {\cal S} = -2 \left[ \alpha \sum_{i=1}^N w_i \ln L_i
    - (\alpha + 1) \ln L^* \sum_{i=1}^N w_i - {1 \over {L^*}}
      \sum_{i=1}^N w_i L_i - \sum_{i=1}^N w_i \ln
      \Gamma\left(\alpha+1, {{L_{min}(z_i)}\over{L^*}} \right)
       \right]    \eqno(5) $$
Loveday et al. (1992) have used eq. (2) to compute the luminosity
function for the galaxies in their sample, both when they considered
all galaxies and galaxies divided in sub--groups, as a function of the
morphological type. Since, as mentioned in the Introduction, their
morphological
classification of early--type galaxies is biased at faint apparent magnitudes,
a weight $w_i(m_i)$ should be associated to each galaxy
and the use of eq. (2) for determining the luminosity function of early--type
galaxies is not correct anymore. In the following section we will quantify,
through simulated samples, the differences between the results obtained
through the use of eq. (3) and (5).
\bigskip
\parn
{\bf 3. Results}
\medskip
\par
In order to estimate in a quantitative way
the error introduced applying eq. (3) instead of eq. (5)
to an incomplete sample, we have used two types of random simulations.
\par
{\bf 1)} We have randomly distributed 5 millions of points following a
Schechter luminosity function with parameters $\alpha=-1.10$ and $M^* = -19.50$
(which are typical values for the galaxy luminosity function),
obtaining a sample of $\sim 11000$ galaxies with $m_{lim}=17.15$.
Then we have introduced in this
sample an incompleteness function, depending on the apparent magnitude of the
galaxies in the form of
$$ f(m)= a (m - m_o)                     \eqno(8) $$
and for each $m$ we have randomly eliminated from the sample a fraction $f(m)$
of galaxies. We assumed that $f(m)=0$ for $m<m_o$, {\it i.e.} for galaxies
brighter than $m_o$ the sample is complete.
We have chosen $m_o =16$, assuming that for galaxies brighter than $16^{th}$
magnitude the morphological classification is relatively easy, and we have used
various values for $a$, corresponding to different values of $<V/V_{max}>$
and, therefore, to different levels of incompleteness. Then we have
computed the parameters of the luminosity function for these samples using
both eq. (3) and eq. (5).
\par
Table 1 lists the derived parameters for three representative cases: column
(1) gives the adopted
incompleteness function, column (2) the $<V/V_{max}>$ of the sample and
column (3) the number of galaxies; the parameters $\alpha$ and $M^*$ derived
from eq. (3) and (5) are listed in columns (4) and (5), (6) and (7),
respectively. From this table it is clear that, as $a$ increases, the use of
eq. (3) produces flatter slopes and brighter $M^*$ with respect to the
real parameters. In all these cases the derived parameters are not compatible
with the real ones, as shown by the confidence ellipses in Fig. 1.
But if we use eq. (5), which takes into account the incompleteness function,
we obtain ``corrected" parameters in very good agreement with the real ones.
\par
{\bf 2)} In the previous case we have used a simulation with a large number
of points in order to minimize the effects of statistical fluctuations in the
test of eq. (3) and (5). Now we try to reproduce as much as possible the
characteristics of the early--type galaxy sample of Loveday et al.
($<V/V_{max}>=0.32$).
We have generated 100 random catalogues, distributed following a Schechter
function with the parameters derived by Efstathiou et al. (1988) for
early--type galaxies ($\alpha=-0.48$ and $M^* = -19.37$). Note that these
values are not consistent with the values $\alpha=+0.2$ and $M^*=-19.71$
found by Loveday et al. (see below in Fig. 2). Then we have
applied the incompleteness function $f(m)= 0.8(m-16)$, which gives
$<V/V_{max}>=0.33$ (this value has been computed as the mean of the values
of the $<V/V_{max}>$ derived for each catalogue). The number of object in
each incomplete catalogue is of the same order as the number of galaxies
in the Loveday et al. sample (311 objects). Finally, we have computed
$\alpha$ and $M^*$ applying both eq. (3) and (5). The results are given
in Table 2, whose columns have the same meaning as in Table 1, except that
in this case all the parameters are the mean of the values derived for each
catalogue. In Fig. 2 we show in the $M^* - \alpha$ plane
the parameter pairs found for each incomplete catalogue: open and solid circles
refer to parameters derived with eq. (3) and (5) respectively, while the star
indicates the location of the input parameters and the cross represents the
Loveday et al. parameters. The dashed and dotted curves are the $1\sigma$
confidence ellipses of the parameters as derived by Efstathiou et al. and
Loveday et al., respectively. The figure shows that there is a clear separation
between the two sets of points: while the solid circles are very well
consistent with the input parameters (star), the open circles are displaced
toward a flatter slope $\alpha$ and a brighter $M^*$.
\medskip
In conclusion, our simulations suggest that, qualitatively, the correction for
incompleteness should move the $(\alpha, M^*)$ parameters of Loveday et al.
toward those determined by Efstathiou et al.. Taken at face value, the shift
between uncorrected and corrected values as resulting from the simulations
is only about half of what is needed to obtain a perfect agreement between
the results for the two samples. With such a shift, however, the two confidence
ellipses would overlap somewhat at the $1\sigma$ level, thus becoming
reasonably consistent with each other. Moreover, it is important to stress
that in our simulations we were forced to use arbitrary incompleteness
function. A detailed knowledge of the functional form of the incompleteness of
the Loveday et al. sample should allow to fully assess the consistency between
the luminosity functions which can be derived from the two samples.
\vfill\eject
{\centerline{\bf References}}
\medskip

\ref Binggeli, B., Sandage, A., Tammann, G.A., 1988, \araa 26 509|

\ref Dressler, A., 1980, \apj 236 351|

\ref Efstathiou, G., Ellis, R.S., Peterson, B.A., 1988, \mnras 232 431|

\ref Loveday, J., Peterson, B.A., Efstathiou, G., Maddox, S.J., 1992,
     \apj 390 338|

\ref Schechter, P., 1976, \apj 203 297|

\ref Sandage, A., Tammann, G.A., Yahil, A., 1979, \apj 232 352|

\vfill\eject
{\centerline{\bf Figure Captions}}
\medskip
\parn
{\bf Figure 1):}
\parn
Confidence ellipses at $1\sigma$ level for the parameters listed in Table 1,
referred to the complete and the three incomplete samples of case 1).
The parameters derived for the incomplete samples are clearly not consistent
with the real ones.
\parn
{\bf Figure 2):}
\parn
Parameter pairs ($M^* - \alpha$) found for each incomplete catalogue of case
2): open and solid circles refer to parameters derived with eq. (3) and (5)
respectively. There is a clear separation between the two sets of points, being
the solid circles very well consistent with the input parameters (star).
The star (and the dashed ellipse) and the cross (and the dotted ellipse) refer
to the parameters derived by Efstathiou et al. and Loveday et al.,
respectively.
\parn

\vfill\eject
\bye